# Influence of temperature on the strain rate sensitivity and deformation mechanisms of nanotwinned Cu

L. W. Yang[1], C. Y. Wang[1], M. A. Monclús[1], L. Lu[2], J. M. Molina-Aldareguía[1,*], J. LLorca[1,3,*]

[1]IMDEA Materials Institute, C/Eric Kandel 2, 28906 –Getafe, Madrid, Spain.
[2]Shenyang National Laboratory for Materials Science, Institute of Metal Research, Chinese Academy of Sciences, Shenyang 110016, P R China
[3]Department of Materials Science, Polytechnic University of Madrid. E. T. S. de Ingenieros de Caminos, 28040 Madrid, Spain
*Corresponding authors: jon.molina@imdea.org, javier.llorca@imdea.org

**Abstract**

The mechanical behavior of nanotwinned Cu was studied through indentation creep and constant strain rate indentation tests from 25°C to 200°C. The results showed an enhanced strain rate sensitivity of nanotwinned Cu with temperature, which was higher than that found in coarse-grained Cu. Transmission electron microscopy revealed the same deformation mechanism in the whole temperature range: confined dislocation slip between coherent twin boundaries and formation of dislocation pile-ups at the coherent twin boundaries. The mechanisms responsible for large increase in strain rate sensitivity of nanotwinned Cu with temperature were discussed to the light of experimental observations.

**Keywords**: Nano twinned Cu; Indentation; High temperature deformation. Strain rate sensitivity

Metals with nanoscale twins have received increasing attention in recent years because of their unique properties [1–5]. In particular, nano-twinned (NT) metals (twin spacing <100 nm) achieve a strength and hardness similar to nanocrystalline (NC) metals (grain size < 100 nm) [6], but are able to maintain substantial ductility [5,7]. In addition, NT metals present outstanding thermal stability, electrical conductivity and fatigue resistance [8]. These differences reflect the particular nature of coherent twin boundaries (CTB), as compared with standard grain boundaries, which lead to the activation of different deformation mechanisms. Obviously, further understanding of the dominant deformation processes under different conditions is necessary to design NT metals with improved properties. However, the current knowledge of the deformation of NT metals is still limited to ambient temperature despite that the plastic deformation of metals is a thermally activated process. This investigation was aimed at exploring the influence of temperature on the mechanical properties and deformation mechanisms of these materials.

Most of the work on NT metals has been focused in Cu. High density of NTs with modulated twin structure can be introduced in Cu by pulsed electrodeposition. NT-Cu processed following this route shows a tensile strength about 10 times higher than that of conventional course-grained (CG) Cu, and an electrical conductivity comparable to that of pure Cu [1]. This combination of properties comes about as a result of the large density of CTBs which block the dislocation slip while presenting extremely low electrical resistivity, that cannot be achieved by other types of grain boundaries. The mechanical behavior of NT-Cu at ambient temperature has been extensively studied by means of uniaxial tension [7,9,10], nanoindentation [9,11–13], micropillar compression tests [14,15] and theoretical models [16]. CTBs and grain boundaries (GBs) act as dislocation sources during deformation and plastic deformation is carried by the formation of dislocation pile-ups along the CTBs [9,17]. Recent studies have revealed that the strain rate sensitivity ($m$) of NT-Cu at ambient temperature and quasi-static loading (strain rate $10^{-4}$ s$^{-1}$~$10^{-1}$ s$^{-1}$) was in the range 0.03~0.08. These values were much higher than those of its coarse-grained (CG) polycrystalline counterparts (0.004~0.007) [13] and contribute to delay necking and increase the ductility during tensile deformation.

In this investigation, the deformation mechanisms of NT-Cu were explored in the temperature range 25 ºC to 200 ºC by means of creep and constant strain rate indentation tests. The results of the mechanical tests (in terms of the strain rate sensitivity and activation volume) were completed with the help of transmission electron microscopy (TEM) observation to ascertain the dominant mechanisms of plastic deformation as a function of temperature.

High-purity Cu (99.99 wt%) sheets with nanoscale growth twins were synthesized by means of direct-current electro-deposition from an electrolyte of $CuSO_4$. More details about the deposition parameters can be found in [18]. All the deposition parameters (including temperature, pH, solution volume, current density, etc.) were kept almost constant during electrodeposition to ensure the homogeneity and consistency of the microstructure of the different sheets. The NT-Cu sheet was deposited on a Ni substrate. The final sheet thickness was > 1.5 mm. A CG Cu with similar grain size prepared by the same electrolyte using DC electrodeposition was also studied for comparison [19]. The initial dislocation density of the CG-Cu was in the range $10^{10}$ to $10^{12}$ m$^{-2}$.

The textures of NT-Cu and CG-Cu sheets were analyzed by means of electron backscatter diffraction (EBSD), using a field-emission gun SEM (Helios Nanolab 600i, FEI) equipped with an HKL EBSD system. The average grain size was determined by the line intercept method from the EBSD micrographs. TEM was used to analyze the microstructure of twins and twin boundaries and their interaction with dislocations. A trenching-and-lifting-out focused ion beam technique was adopted to extract lamellae directly from the indentation imprints, which was subsequently thinned to approximately <100 nm for electron transparency. TEM observation was carried out in two-beam diffraction mode [20], employing different diffraction vectors, $g$, using the $g \cdot b = 0$ invisibility criterion, where $b$ stands for the direction of the Burgers vector.

The mechanical properties of NT-Cu and CG-Cu samples were measured by means of indentation creep tests in a NanoTest$^{TM}$ platform III (Micro Materials, Wrexham, UK). This platform uses an independent tip and sample heating system that is designed to maintain the temperature within ±0.05°C, which is the best strategy to achieve thermal equilibrium during indentation in order to minimize thermal drift. Load was applied with a diamond Berkovich indenter. It was initially increased up to 50 mN in 10 s in the nanoindentation creep tests and was maintained constant during 400 s followed by unloading at the rate of 5 mN/s. The evolution of the indentation depth, $h$, *vs.* time, $t$, was recorded during the holding time. Tests were carried out at 25 °C, 50 °C, 100 °C, 150 °C and 200 °C, respectively. At each temperature, thermal drift was carefully minimized (<0.1 nm/s) to equilibrate the temperatures of both the indenter and the sample. At least 5 tests were performed at each temperature and the results presented below for each temperature are the average value and the standard deviation of these tests.

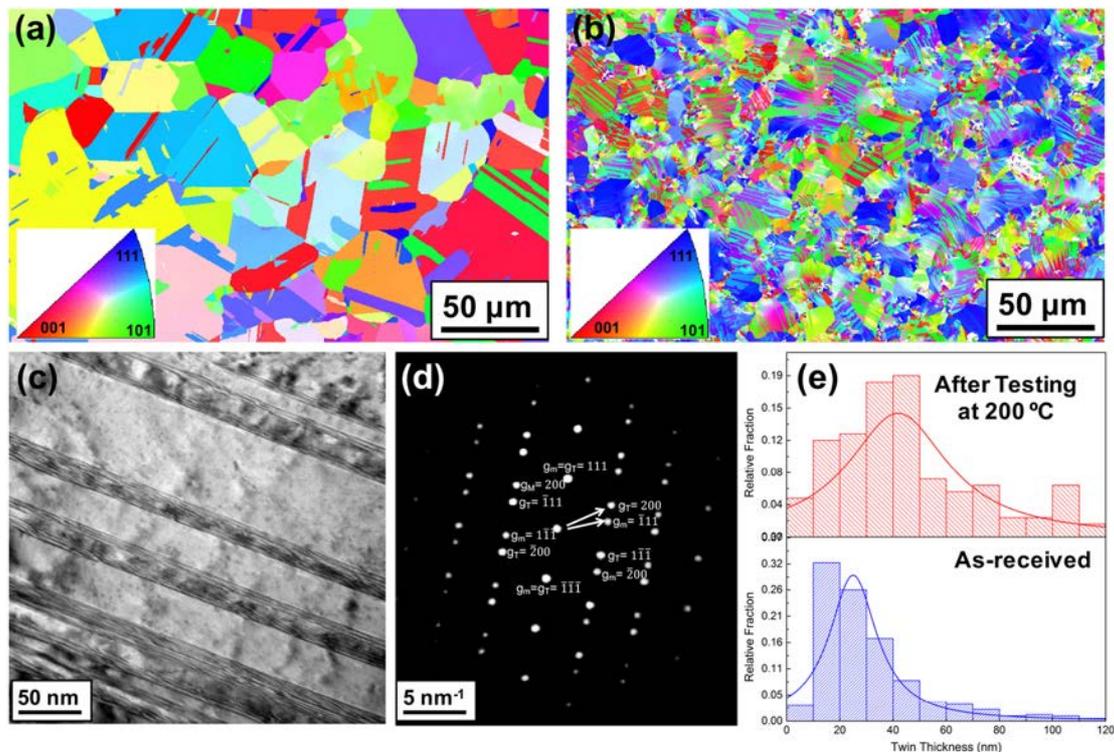

Fig. 1. (a) EBSD map of CG-Cu. (b) *Idem* of NT-Cu. (c) High resolution TEM of twin structure within a grain of NT-Cu. (d) Selected-area diffraction pattern of the NT-Cu in the zone axis <01$\bar{1}$>. (e) Histograms of twin thickness in NT-Cu in the as-received condition and after testing at 200 °C.

The EBSD maps of the as-received CG-Cu and NT-Cu are shown in Figs. 1(a) and (b), respectively. The average grain size of the CG-Cu was 22 ± 4 μm with random texture and only a few twins were found within each grain. The average grain size of the NT-Cu was 14 ± 2 μm and the microstructure presented a strong (111) texture, which may result in smaller grain misorientation, and, therefore, reduces the driving force for grain growth at high temperatures [21, 22]. High density of twins was found in most grains of the NT-Cu, leading to nm-thick twin/matrix lamellar structures, as shown in Fig. 1(c). The diffraction spots of the twins are clearly seen in the selected-area diffraction pattern of NT-Cu in the zone axes <01$\bar{1}$> in Fig. 1(d). Only a few dislocations pinned at CTBs were observed in the as-received NT-Cu. The nanotwinned structure was retained after thermal exposure up to 200 °C. Fig. 1(e) plots the statistical distribution of twin lamella thickness in the as-received samples and after testing at 200 °C. The average twin thickness in the as-received sample was ≈25 nm and increased up to ≈40 nm after thermal exposure. It has been reported that the CTBs are more thermally stable than the grain boundaries due to their low coherent energy (24-39 mJ/m$^2$) [22, 23] and, thus, the thermal exposure during testing up to 200 °C did not alter the NT structure significantly.

The indentation creep data were analyzed following a standard protocol [24]. The increase in the indentation depth, $h$-$h_0$ (measured from the initial indentation depth at the maximum load of 50 mN, $h_0$) as a function of time, $t$, was approximated by an empirical law

$$h(t) - h_0 = x(t - t_0)^y + z(t - t_0) \qquad (1)$$

where $h_0$, $x$, $y$ and $z$ stand for fitting parameters and $t_0$ (= 10 s) is the time necessary to attain the maximum load. Eq. (1) was able to reproduce very accurately the indentation creep curves for both NT-Cu and CG-Cu. The indentation strain rate was thus given by

$$\dot{\varepsilon} = \frac{1}{h}\left(\frac{dh}{dt}\right) \qquad (2)$$

In addition, an equivalent hardness, $H$, was calculated by dividing the peak load (50 mN) by the apparent contact area, $A_{ap}$, that was estimated from the indentation depth $h$ using actual tip area function. The strain rate sensitivity, $m$, was finally obtained by the slope of the double logarithmic plot of $H$-$\dot{\varepsilon}$ according to

$$m = \frac{\partial \ln(H)}{\partial \ln(\dot{\varepsilon})} \qquad (3)$$

The strain rate sensitivities determined from the indentation creep tests were verified through another independent indentation-based method, which relies on indenting the samples at a constant strain rate (CSR). The strain rate sensitivity was also determined by fitting the logarithmic $H$-$\dot{\varepsilon}$ curves. The tests performed in the NanoTest™ platform from 25°C to 200°C. CSR tests were carried out at three different strain rates (0.002 s$^{-1}$, 0.01 s$^{-1}$ and 0.05 s$^{-1}$). The peak load (50 mN) was the same in the CSR tests and in the creep tests to achieve comparable testing volumes, corresponding to an initial indentation depth of ≈1 μm for NT-Cu and ≈1.5 μm for CG-Cu (Figs. 2b and 2d).

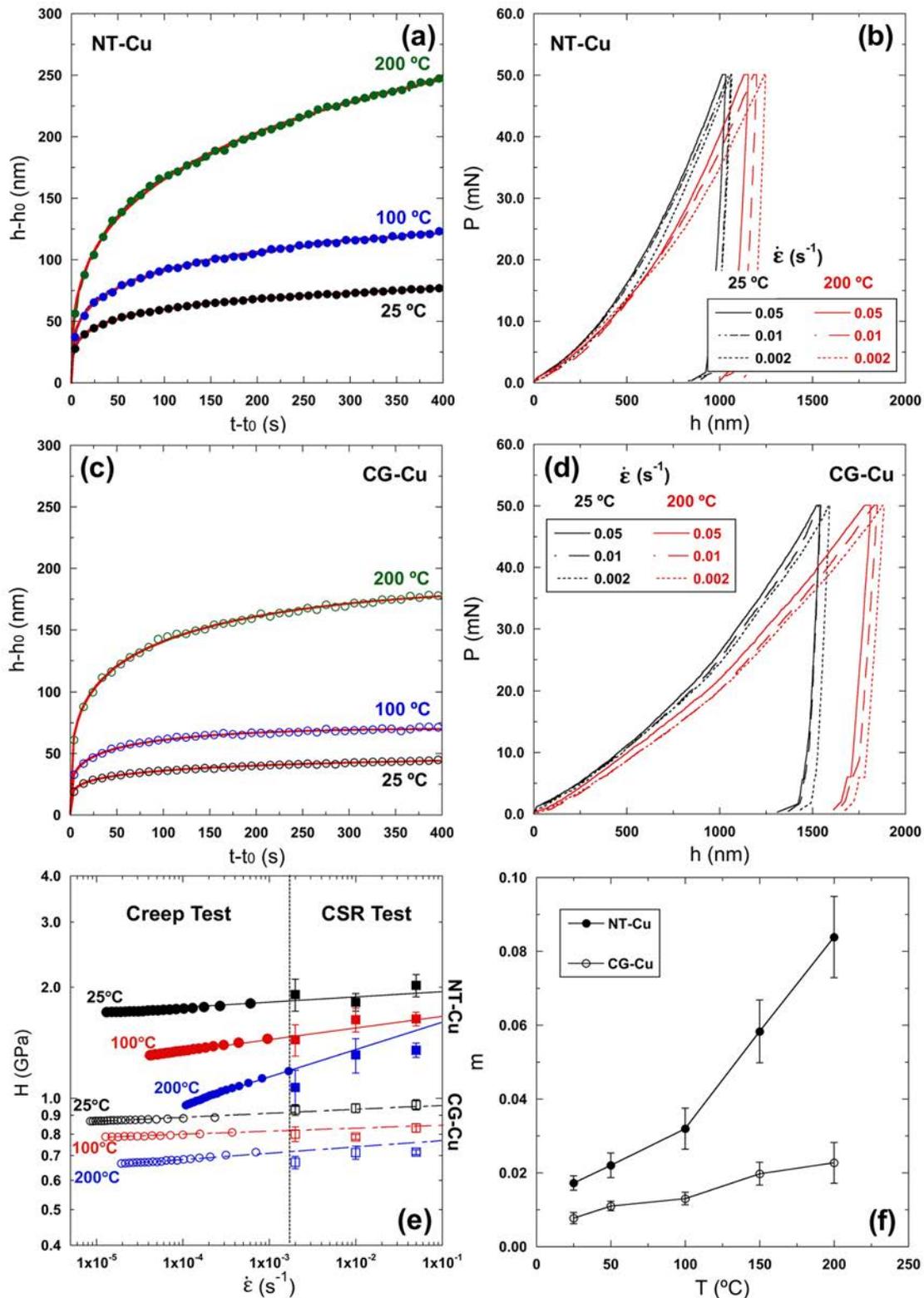

Fig. 2. (a) Representative indentation creep curves of NT-Cu at different temperatures. (b) Representative CSR indentation curves of NT-Cu at 25 ºC and 200 ºC. (c) *Idem* as (a) of CG-Cu. (d) *Idem* as (b) of CG-Cu. (e) *H-ἐ* in bilogarithmic coordinates for CG-

Cu and NT-Cu at different temperatures. (f) Evolution of the strain rate sensitivity with temperature CG-Cu and NT-Cu.

Representative indentation creep curves at 25 ºC, 100 ºC and 200 ºC are plotted in Fig. 2(a), showing the evolution of the indentation depth, $h$-$h_0$, with time $t$-$t_0$. The fitting of the experimental data with Eq. (1) is also included for comparison. The load, $P$, vs. indentation depth, $h$, curves corresponding to the CSR tests of NT-Cu at 25 ºC and 200 °C are plotted in Fig. 1(b) for different strain rates. The $H$-$\dot{\varepsilon}$ curves of the NT-Cu and CG-Cu at 25 °C, 100 °C and 200 °C were computed over the entire duration of the creep tests (400 s). They are plotted in double logarithmic coordinates in Fig. 2(e). They include data obtained independently by indentation creep and CSR indentation tests. The strain rates of the CSR tests were higher than those obtained in the indentation creep tests. It was found that the data for both materials follow Eq. (3) in a very wide range of strain rates ($10^{-5}$ s$^{-1}$ to $10^{-1}$ s$^{-1}$) and temperatures (from 25 ºC to 200 ºC). This result evidences the robustness of both testing methods to determine the strain rate sensitivity of Cu up to 200 °C.

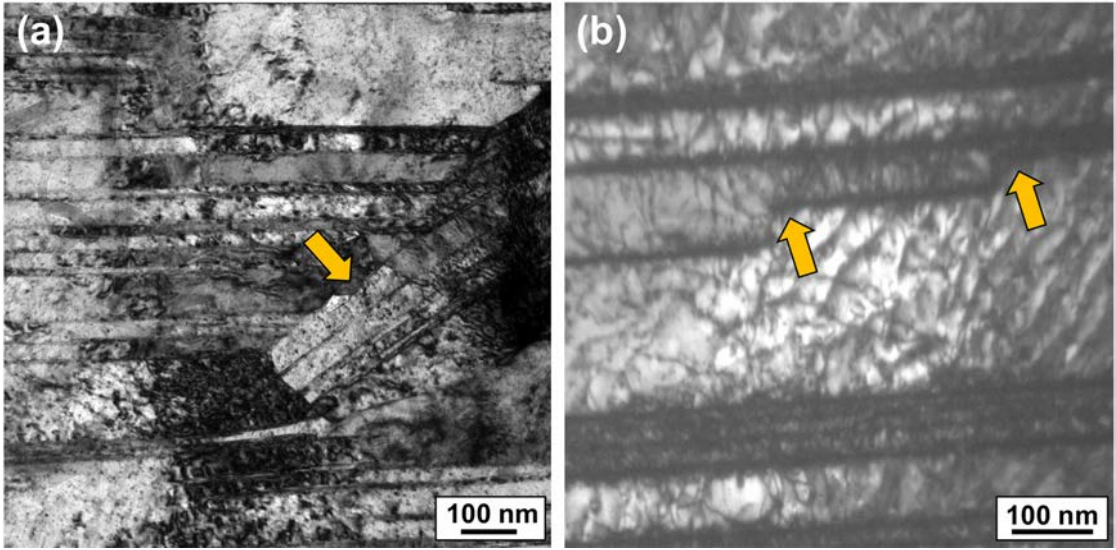

Fig. 3. TEM images of NT-Cu tested at (a) 25 °C and (b) 200 °C showing a cross-section beneath the indentation imprints. The yellow arrows indicate the displacement of CTBs. The zone axis of the images was [01$\bar{1}$]. The lamellae were extracted from the plastically deformed region below the indenter.

The strain rate sensitivities (given by the slope of ln($H$) – ln($\dot{\varepsilon}$) data) of the NT-Cu and CG-Cu are plotted in Fig. 2(f). They increased with temperature in both cases but NT-Cu presented higher strain-rate sensitivity than the CG-Cu in the whole temperature range and the differences increased rapidly with temperature. The strain rate sensitivities of the NT-Cu (≈0.017) and CG-Cu (≈0.008) at ambient temperature were consistent with those found in the literature [25, 26]. For the CG-Cu, increasing temperature from 25°C to 200°C yielded a slight increase in $m$ from ≈0.008 to ≈0.023. This is typical of CG polycrystalline fcc metals, whose plastic deformation at low homologous temperatures ($T/T_m$ < 0.35) is mainly controlled by dislocation-dislocation interactions (forest hardening) and dislocation-grain boundary interactions [27]. The strain rate sensitivity of NT-Cu was consistently lower than that of ultra-fine-grain (UFG)-Cu with an average grain size ≈300 nm in the entire temperature range [28].

However, the increase in the strain rate sensitivity with temperature was much more pronounced for NT-Cu: from ≈0.017 at 25°C to ≈0.084 at 200°C, as opposed to UFG-Cu, from ≈0.20 at 25°C to ≈0.26 at 200°C. In summary, the strain rate sensitivity of NT-Cu increased by 400% between 25 °C and 200 °C and by 190% in CG-Cu and by 30% in UFG-Cu in the same temperature range.

TEM micrographs of lamellae extracted from the indentation imprints of NT-Cu are shown in Fig. 3. The nanotwinned structure was maintained after the plastic deformation at both 25 °C and 200 °C beneath the sharp indenter. Large dislocation densities were found within nanotwin lamellae, particularly in thinner ones. Most dislocations were accumulated and pinned at the CTBs at both 25 °C (Fig. 3a) and 200 °C (Fig. 3b), leading to the high hardness of the NT-Cu. In addition, displacement of CTBs (marked with arrows) and formation of jogs and steps in the CTBs were also observed at both temperatures. It has been assumed that these defects in the CTBs act as dislocation nucleation sites and provide the large dislocation density (which was not found in the as-received material, Fig. 1c) necessary to maintain the plastic flow [9]. These observations indicate that the deformation mechanisms of NT-Cu did not change in the temperature range 25 °C to 200 °C, i.e. nucleation of dislocation at CTBs defects followed by confined layer slip of the dislocations between CTB and by the pile-up of the dislocations at the CTBs.

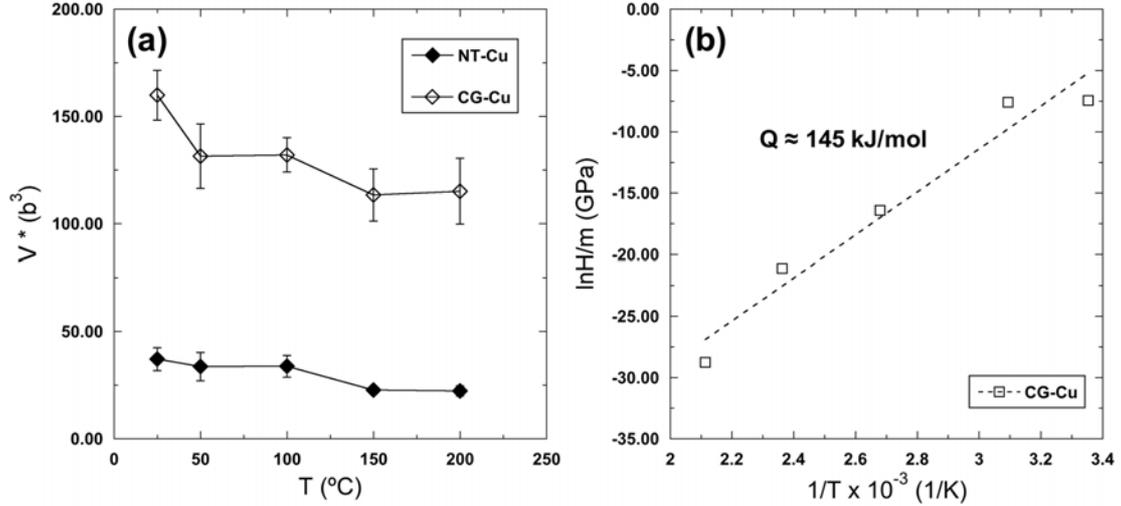

Fig. 4. (a) Activation volume, $V^*$, of NT-Cu and CG-Cu as a function of temperature. (b) Activation energy, $Q$, for the plastic deformation of CG-Cu in the temperature range 25 °C~200 °C.

Assuming that the flow stress is equivalent to 1/3 of the apparent hardness $H$, which is reasonable for metals [29], the activation volume ($V^*$) can be related to the strain rate sensitivity ($m$) according to

$$V^* = \frac{3\sqrt{3}kT}{Hm} \quad (4)$$

where $k$ is the Boltzmann constant and $T$ the absolute temperature. The activation volumes of NT-Cu and CG-Cu in the range 25 °C~200°C are plotted in Fig. 4(a). The activation volume of CG-Cu only decreased slightly from $160b^3 - 120b^3$ ($b$ is the burgers vector) in this temperature range, indicating that the dominant deformation mechanism did not change [30]. These values are smaller than the conventional activation volumes of CG-Cu (few hundred to one thousand $b^3$) when the dominant hardening mechanism is the interaction of forest dislocations [30-31] but the differences may be attributed to the fact that the steady-state conditions are not reached during the creep indentation tests. Grain boundary diffusion is not expected to be critical in this temperature range because of the large grain size ($\approx$ 50 μm) and the limited temperature ($T/T_m < 0.35$).

The activation volume of NT-Cu at 25 °C was $\approx 40b^3$, which is comparable to those reported in the literature for NT-Cu [9,13] as well as for UFG-Cu [28] and UFG-Ni [32] with average grain sizes of $\approx$ 300 nm and $\approx$ 30 nm, respectively. The smaller activation volumes of NT-Cu and UFG-Cu and UFG-Ni (as compared with CG-Cu) reflect the differences in the dominant deformation mechanisms in these materials. They are related to the interaction of confined dislocations with CTBs within very narrow twin lamellae in NT-Cu and to the interaction of dislocations with GB in UFG-Cu and Ni.

The activation volume of NT-Cu decreased to $\approx 20b^3$ at 200 °C and this behavior is similar to the one reported in UFG-Ni [29], while the activation volume of UFG-Cu increased from $\approx 40b^3$ at ambient temperature to $\approx 60b^3$ at 200 °C [28]. The variation of the activation volumes with temperature in NT-Cu and UFG-Ni and Cu was more noticeable (in relative terms) than that in the case of CG-Cu but it does not indicate a change in the deformation mechanisms, which was neither found in the TEM observations of NT-Cu nor in the investigations of UFG-Ni and Cu.

The activation energy, $Q$, of the dominant deformation processes CG-Cu was finally estimated according to

$$\dot{\varepsilon} = CH^{1/m} \exp(-\frac{Q}{RT}) \qquad (5)$$

where $C$ is a pre-factor. This equation can be written of the form

$$\frac{1}{m} \ln H = \frac{Q}{RT} + \ln \dot{\varepsilon} - \ln C \qquad (6)$$

and the activation energy can be obtained from the slope $\frac{1}{m} lnH$ vs. $(1/T)$ assuming that the strain rate sensitivity exponent $m$ is independent of the temperature. This hypothesis is supported by the experimental data in Fig. 2f for CG-Cu but not for NT-Cu and the activation energy could not be determined for the latter. The corresponding results for $\dot{\varepsilon} = 10^{-4}$ s$^{-1}$ are plotted in Fig. 4d for CG-Cu assuming an average value of the strain rate sensitivity in the range 25 °C-200 °C. The activation energy for CG-Cu ($\approx$145 KJ/mol) is similar to the one reported by pipe diffusion along dislocations in Cu in the temperature range 210°C to 350°C [33] and is compatible with a rate controlling mechanism which depends on this phenomenon. The deformation mechanisms in NT-Cu did not change with temperature in the range 25 °C to 200 °C according to the TEM observations in Fig. 3, but strain rate sensitivity showed a dramatic increase in the same temperature range. This variation in the strain rate sensitivity can be accounted for a change in the rate limiting mechanism during deformation. Plastic flow in NT-Cu involves the nucleation of dislocations from CTB, the propagation of dislocations between these boundaries by confined layer slip and the pile-up of the dislocations against the CTB and may also be influenced by twin boundary migration [9, 16-17]. The mechanical behavior at 25 ºC was rather strain rate insensitive and this fact, together with the high hardness level and the activation volume of ≈40$b^3$ is compatible with a rate-limiting deformation mechanism controlled by confined layer slip [12]. The activation volume was reduced to ≈20$b^3$ at 200 ºC but this value is too large to assume that dislocation nucleation at the CTB is the controlling mechanism. Moreover, the slight increment in the average twin thickness with temperature (Fig. 1e) while the activation volume is reduced does not support that twin boundary migration is responsible for the high temperature behavior. Thus, dislocation climb at the dislocation pile-ups associated with the CTBs seems to be the most likely rate-limiting mechanism at high temperature, which is responsible for the enhanced strain rate sensitivity at 200 ºC. Moreover, the activation energy for this process will be reduced by the high dislocation density between the CTBs, which enhances the diffusion along dislocations.

In summary, the mechanical behavior of NT-Cu was studied independently by means of indentation creep and constant strain rate tests from 25°C to 200°C. The results showed that NT-Cu presented higher strain-rate sensitivity than the CG-Cu in the whole temperature range and the differences increased rapidly with temperature (from a factor of ≈2 at 25 ºC to ≈4 at 200 ºC). Transmission electron microscopy analysis showed the twinned structure was maintained under the indenter during the high temperature tests and revealed similar deformation mechanisms for NT-Cu in the whole temperature range: nucleation of dislocations at CTB, followed by confined slip of dislocations between CTB and pile-up of dislocations at the CTBs. The activation volume for NT-Cu (40$b^3$–20$b^3$) was compatible with this mechanism. The creep activation energy for CG-Cu was in good agreement with that reported for diffusion along dislocations but it could not be determined for NT-Cu because of the large increase in strain rate sensitivity with temperature. This behavior was attributed to a change in the rate-limiting mechanism during deformation from confined layer slip at 25 ºC to dislocation climb at the CTBs at 200 ºC.

**Acknowledgements**

This investigation was supported by the European Research Council (ERC) under the European Union's Horizon 2020 research and innovation programme (Advanced Grant VIRMETAL, grant agreement No. 669141. L. W. Yang and C. Y. Wang acknowledges the financial support from China Scholarship Council (grant numbers: 201306110007 and 201406290011).

**References**

[1]   L. Lu, Y. Shen, X. Chem, L. Qian, K. Lu, Science. 304 (2004) 422–426.
[2]   T. Zhu, J. Li, A. Samanta, H.G. Kim, S. Suresh, Proc. Natl. Acad. Sci. 104 (2007) 3031–3036.
[3]   Y. Kulkarni, R.J. Asaro, Acta Mater. 57 (2009) 4835–4844.
[4]   J. Bezares, S. Jiao, Y. Liu, D. Bufford, L. Lu, X. Zhang, Y. Kulkarni, R.J. Asaro, Acta Mater. 60 (2012) 4623–4635.
[5]   P. Zhou, Z.Y. Liang, R.D. Liu, M.X. Huang, Acta Mater. 111 (2016) 96–107.
[6]   Y.M. Wang, A. V. Hamza, E. Ma, Acta Mater. 54 (2006) 2715–2726.

[7]    M. Dao, L. Lu, Y.F. Shen, S. Suresh, Acta Mater. 54 (2006) 5421–5432.
[8]    Q. Pan, H. Zhou, Q. Lu, H. Gao, L. Lu, Nature. 551 (2017) 214.
[9]    L. Lu, R. Schwaiger, Z.W. Shan, M. Dao, K. Lu, S. Suresh, Acta Mater. 53 (2005) 2169–2179.
[10]   Y.F. Shen, L. Lu, M. Dao, S. Suresh, Scr. Mater. 55 (2006) 319–322.
[11]   J. Chen, L. Lu, K. Lu, Scr. Mater. 54 (2006) 1913–1918.
[12]   J.C. Ye, Y.M. Wang, T.W. Barbee, A. V. Hamza, Appl. Phys. Lett. 100 (2012) 261912.
[13]   I.-C. Choi, Y.-J. Kim, Y.M. Wang, U. Ramamurty, J.-I. Jang, Acta Mater. 61 (2013) 7313–7323.
[14]   M. Mieszala, G. Guillonneau, M. Hasegawa, R. Raghavan, J.M. Wheeler, S. Mischler, J. Michler, L. Philippe, Nanoscale. 35 (2016) 15999-16004.
[15]   A.T. Jennings, J. Li, J.R. Greer, Acta Mater. 59 (2011) 5627–5637.
[16]   X. Xiao, D. Song, H. Chu, J. Xue, H. Duan, Int. J. Plasticity 74 (2015) 110.
[17]   F. Sansoz, K. Lu, T. Zhu, A. Misra, MRS Bull. 41 (2016) 292–297.
[18]   Z. You, X. Li, L. Gui, Q. Lu, T. Zhu, H. Gao, L. Lu, Acta Mater. 61 (2013) 217–227.
[19]   Z. S. You, L. Lu, K. Lu. Acta Materialia, 59 (2011) 6927-6937.
[20]   Q. Lu, Z. You, X. Huang, N. Hansen, L. Lu, Acta Mater. 127 (2017) 85–97.
[21]   Y. Zhao, T.A. Furnish, M.E. Kassner, A.M. Hodge, J. Mater. Res. 27 (2012) 3049-3057.
[22]   O. Anderoglu, A. Misra, H. Wang, X. Zhang, J. Appl. Phys. 103 (2008) 094322.
[23]   X. Zhang, A. Misra, Scr. Mater. 66 (2012) 860–865.
[24]   R. Goodall, T.W. Clyne, Acta Mater. 54 (2006) 5489–5499.
[25]   M.A. Meyers, A. Mishra, D.J. Benson, Prog. Mater. Sci. 51 (2006) 427–556.
[26]   Q. Wei, S. Cheng, K.T. Ramesh, E. Ma, Mater. Sci. Eng. A. 381 (2004) 71–79.
[27]   U.F. Kocks, H. Mecking, Prog. Mater. Sci. 48 (2003) 171–273.
[28]   T. Suo, L. Ming, F. Zhao, Y. Li, X. Fan, Int. J. Appl. Mech. 5 (2013) 1350016.
[29]   D. Tabor, The Hardness of Metals, Oxford University Press, 1951.
[30]   H. Conrad, Mater. Sci. Engng. A341 (2003) 216-228.
[31]   R.P. Carreker, W.R. Hibbard, Acta Metall. 1 (1953) 654–663.
[32]   Y.M. Wang, A.V. Hamza, E. Ma, Acta Mater. 54 (2006) 2715.
[33]   D.B. Butrymowicz, J.R. Manning, M.E. Read, J. Phys. Chem. Ref. Data. 2 (1973) 643–656.